# Fabrication of uniform half-shell magnetic nanoparticles and microspheres with applications as magnetically modulated optical nanoprobes

Brandon H. McNaughton[1,2], Vladimir Stoica[2], Jeffrey N. Anker[1,2], Roy Clarke[2], and Raoul Kopelman[1,2,3]

1. The University of Michigan, Department of Chemistry, Ann Arbor, MI 48109-1055
2. The University of Michigan, Applied Physics Program, Ann Arbor, MI 48109
3. Electronic mail: kopelman@umich.edu

(Dated June 15, 2005)

Magnetically Modulated Optical Nanoprobes (MagMOONs) are magnetic particles that indicate their angular orientation by emitting varying intensities of light for different orientations and have shown promise for a variety of applications. In this letter we describe a new method to fabricate uniform magnetic half-shell particles that can be used as MagMOONs. Cobalt was deposited onto commercially made polystyrene nanospheres and microspheres, using ultrahigh vacuum vapor deposition, producing particles with uniform size, shape and magnetic content. Additionally, the coercivity of the cobalt deposited on the nanospheres was enhanced compared to its bulk value.

## I. INTRODUCTION

There are two principal approaches to fabrication of nanodevices: top-down, and bottom-up. The top-down approach builds nanodevices by controlled deposition of materials onto substrates[1]. The bottom-up approach builds nanodevices by assembling monomers, dyes, and crystals into nanoparticles by wet-chemistry methods[2,3]. Both approaches have their advantages: the top-down approach allows fine and systematic control of the material and direct electrical interfacing; while the bottom-up approach allows rapid production[4], flexibility of design[2], and creation of devices for *in situ* control and measurement[5]. Combining bottom-up synthesis of nanospheres and microspheres with top-down methods for controlled deposition of magnetic and semiconductor materials, we produced hybrid particles with well controlled magnetic, optical, electronic, and chemical properties.

Magnetic nanoparticles and microspheres have shown promise in a wide range of bio-medical applications[6]. Generally, these magnetic particles are synthesized in bottom up techniques[3-6]. However, limited uniformity often results from synthesis involving magnetic materials and this can affect the magnetic responsiveness of the particles[7]. Indeed Häfeli et al. demonstrated that many commercially produced particles have large variations (33%-80% in a viscous solution) in their magnetophoretic mobility[7]. Variations in magnetic responsiveness reduce the reliability of magnetic probes, especially for single particle force and torque measurements on the viscoelastic properties of biomolecules and cells. Such measurements have required ensemble averaging over many particles or calibration of individual particles. In an effort to produce anisotropic but uniform magnetic particles, we demonstrate an alternative fabrication method: depositing magnetic materials smoothly onto uniformly sized nanospheres and microspheres, supported by a substrate, and then suspending and manipulating the resulting half-shell coated spheres in solution. This produces magnetic half-coated spheres that are uniform in surface roughness, shape, and size.

Other investigators have also coated monodispersed spheres with a variety of materials, but for different purposes than ours. For example, nanosphere lithography (NSL) uses a layer of nanospheres and microspheres as a deposition mask[8]. Generally, the particles are removed from the substrate and discarded, leaving behind hexagonal closed packed interstices of the deposited material on the substrate. In some cases, the deposited material is ferromagnetic[9]. The previously discarded polymer spheres are half-coated and exhibit structural, chemical, magnetic, electric, and optical anisotropies[10-11,12]. Instead of discarding these spheres, we use them to measure and probe various chemical and physical properties. The anisotropies of these half-shell coated spheres, especially the magnetic and optical anisotropies, allow for the creation of particles that have angle dependant photonic intensities. These types of anisotropic particles are known as Magnetically Modulated Optical Nanoprobes (MagMOONs)[10,13-19]. In solution, a MagMOON can be magnetically rotated by an external magnetic field, from the "off" orientation with the metal coated hemisphere in view to the "on" orientation with the uncoated hemisphere in view. This property allows for many interesting applications, namely for photonic background subtraction[10], microrheology[18], and immunoassays[16]. In this letter, we demonstrate the production of uniform half-shell magnetic particles, fabricated using ultrahigh vacuum (UHV) vapor deposition, and their use as MagMOONs. Their characteristics are investigated by using reflection high energy electron diffraction (RHEED), scanning electron microscopy (SEM), magneto optic Kerr effect (MOKE) magnetometry, fluorescent microscopy, and transmission microscopy.

## II. EXPERIMENTAL

MagMOONs were produced by coating nanospheres and microspheres with a half-shell layer of ferromagnetic cobalt, using UHV vapor deposition. Polystyrene spheres of 50 nm, 300 nm, 1 μm, 2 μm, and 3.4 μm in diameter, purchased from Bangs Labs and Spherotech, were coated with different thicknesses of cobalt, 5 nm, 20 nm, and 30 nm. The cobalt was deposited at an evaporation rate of 0.1–0.4 Å/s with a chamber background pressure of $10^{-9}$ torr, providing precise control of the amount of deposited cobalt. In addition, iron, silver, and nickel have also been deposited onto spheres with the system. For comparison purposes, 340 nm carboxyl magnetic nanoparticles, 1.89 μm carboxyl magnetic microspheres, and 4.4 μm diameter fluorescent chromium dioxide microspheres were purchased from Spherotech (Libertyville, IL). As shown in Figure 1, a four-step fabrication process for producing uniform half-shell magnetic particles was followed. We used a glass microscope slide as a substrate to support the spheres, the deposition metal was cobalt, and the spheres were removed from the substrate with a razor blade.

All SEM images were obtained using a Philips XL30 field emission gun. Magnetic hysteresis data were obtained using MOKE magnetometry. RHEED images were taken *in situ* using a kSA 400 system (k-Space

Associates Inc., Ann Arbor, MI). The MagMOONs were observed microscopically with an Olympus IMT-II (Lake Success, NY) inverted fluorescence microscope, and imaged with the Roper Coolsnap ES CCD camera (Roper Scientific). To excite fluorescence, a Lambda DG-4 Xenon Lamp (Sutter Instruments, Novato, CA) was used in combination with an Olympus blue filter cube.

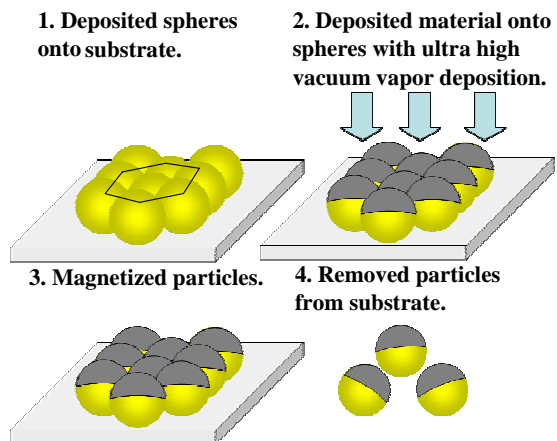

**Figure 1**: Schematic four-step process for fabrication of uniform magnetic particles that can be used as Magnetically Modulated Optical Nanoprobes (MagMOONs), where, in step 1, the spheres form a hexagonal closed packed pattern when forming a monolayer.

The MagMOONs were rotated by an externally rotating permanent cylindrical magnet, magnetized along its diameter, located several centimeters above the sample. The external magnet was connected to a stepper motor and controlled by a function generator. Movies of rotating MagMOONs were acquired and analyzed using Metamorph software (Universal Imaging Corp).

### III. RESULTS & DISCUSSION

RHEED images obtained during cobalt deposition revealed ring-like diffraction patterns, which indicate that the deposited cobalt was polycrystalline. These rings in the RHEED pattern, evident at lower thicknesses, became sharper and more intense as the cobalt thickness was increased, indicating a better coverage of the nano and microspheres.

After the MagMOONs were coated, SEM images were taken, Figure 2. The top row of Figure 2 depicts magnetic half-shell particles fabricated using UHV vapor deposition, where the direction of the arrow indicates increasing size, from 300 nm to 3.4 μm. For comparison, the bottom row of Figure 2 shows similarly sized commercially made magnetic particles. These commercially made particles appear to be rough, irregular in shape (>20% deviation from a perfect circle for 340 nm particles) and have a large variation in size, especially for smaller particles; by comparison, the cobalt coated particles are smooth, have close to spherical shape (<10% deviation from a perfect circle for 300 nm particles), and have a narrower size distribution. To investigate variations in coating uniformity from particle to particle, the 3.4 μm cobalt coated particles were imaged with bright field (transmission) microscopy. On a flat surface the 30 nm of cobalt attenuated light by 10.6%. The images, Figure 3, revealed particles with almost identical transmission profiles and small variations in maximum transmission (6% standard deviation). Large deviations in the amount of magnetic material will affect magnetic responsiveness[7]. In our own experiments, using commercially available 4.4 µm magnetic microspheres for microrheology[18], we observed that measurements varied significantly from sphere to sphere. With uniform magnetic particles, more accurate experiments could be designed, so as to better probe microrheology[18] and molecular interactions[20]. Additionally, this fabrication technique could be used to modify solid state sensors[2] into MagMOONs, by the simple step of depositing a ferromagnetic metal onto the surface of a sensor.

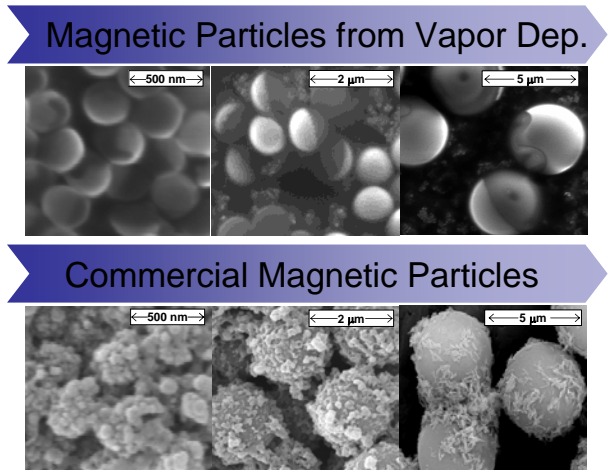

**Figure 2**: Scanning electron microscopy images of various magnetic particles, where the arrow indicates increasing size. Top Row: Magnetic particles fabricated by coating polystyrene spheres that are 300 nm in diameter with a layer of 20 nm of cobalt; 1.0 μm in diameter spheres coated with 30 nm of cobalt; and 3.4 μm spheres coated with 20 nm of cobalt, using ultra high vacuum (UHV) vapor deposition. Bottom Row: Commercially made magnetic particles that are 330 nm, 1.89 μm, and 4.4 μm in diameter (Spherotech, Inc.).

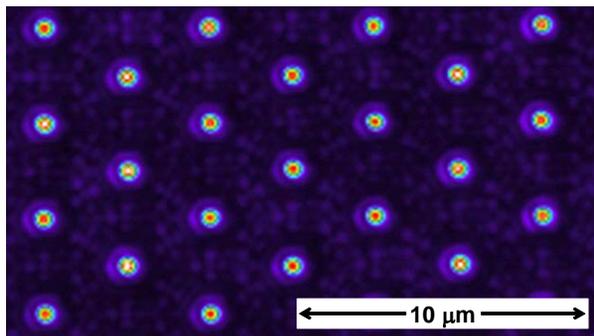

**Figure 3**: Transmission image of a monolayer of 3.4 µm polystyrene microspheres, coated with 30 nm of cobalt, where the pseudocolor scales from blue, representing least transmission, to white, representing highest transmission.

Also evident from the SEM images are arch-like structures in the deposited metal of the 300 nm and the 3.4 µm cobalt-coated spheres, top row of Figure 2. Our current hypothesis is that these arches resulted from either allowing the substrate temperature to exceed the glass transition temperature of polystyrene (≈95°C) or from the anisotropy of deposition that results when the cobalt atomic beam flux forms an angle with the substrate normal. If the substrate temperature was above the glass transition temperature of polystyrene, then the particles could fuse together, and when broken apart leave an arch-like structure on the surface of the sphere. The eye-shaped patterns inside the arches provides evidence for melting and sticking together for the 3.4 µm spheres, top right of Figure 2. Alternatively, if coating occurs at a significant angle normal to the substrate, in our case ~40°, then the arches could form from shadowing effects. This may be the case for the 300 nm particles, top left of Figure 2, where the spheres maintained their spherical shape throughout the deposition. The presence of these arches could affect the magnetic properties of the cobalt film. Additionally, the lightning rod effect may create large electromagnetic field enhancements at the tips with applications for SERS (surface enhanced Raman spectroscopy) and for non-linear optical effects, similar to enhancements seen with prism shaped particles[8] and nanocrescents[12].

To further characterize the magnetic half-shell particles fabricated by UHV vapor deposition, MOKE magnetometry in longitudinal geometry was performed on 50 nm spheres coated with a 20 nm layer of cobalt. The presence of these nanospheres enhanced the coercivity of the planar cobalt layer by more than a factor of three, compared to bulk values, Figure 4a. A similar effect for polar MOKE magnetometry, studied by Albrecht et al.[21], has also been observed in magnetic multilayers of cobalt and palladium grown on polystyrene nanospheres, which may allow for applications in computer memory and magnetic storage. The magnetic hysteresis curves for the 50 nm spheres with the magnetic field applied parallel to the substrate, Figure 4a, have a square-like shape, indicating that the easy axis of magnetization is parallel with the substrate surface as well as with the nanosphere surface.

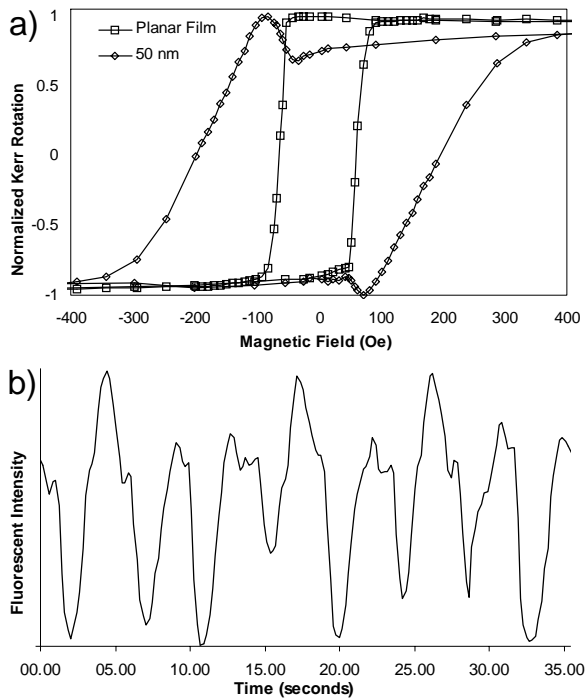

**Figure 4**: Characterization of uniform magnetic particles using a) magnetic hysteresis curves obtained by magneto optic Kerr effect (MOKE) magnetometry on 50 nm polystyrene spheres, coated with 20 nm of cobalt. The asymmetries and kinks near zero-field are due to the vectorial nature of MOKE, which simultaneously probes the in-plane and out-of-plane components of the magnetization. b) Time series of the fluorescence intensity from a 3.4 μm cobalt coated MagMOON rotated by a 0.25Hz magnetic driving field.

In addition to forming magnetic half-shell particles on a plane, we removed them from the glass substrate, suspended them into solution, and rotated them with an external magnetic field. Figure 4b shows the time series for a 3.4 μm fluorescent polystyrene microsphere with a cobalt layer of 20 nm, modulated in an external rotating magnetic field at a driving frequency of 0.25 Hz. Upon demodulation, by subtracting "On" minus "Off," the background can be removed, thus allowing for more sensitive fluorescence measurements. The same process could be utilized with almost any nanosphere or microsphere, thereby allowing nanosensors or microsensors to be modified into MagMOONs.

## IV. CONCLUSION

We have demonstrated that MagMOONs can be fabricated by coating a thin layer of cobalt onto nanospheres and microspheres, using UHV vapor deposition, and that they can be successfully manipulated in solution. The procedure enables fine control over material composition and coating thickness, as well as for layer uniformity. The particles produced are smoothly coated with a reproducible and controllable amount of material, which creates particles limited only by the uniformity of the starting particles. The process works for a wide range of particle sizes, shapes, and compositions, as well as for different material matrixes, providing both a universal method of producing MagMOONs and a method for controlling the particle geometry and the resulting properties; e.g. the coercivity of polycrystalline cobalt was enhanced by the presence of polystyrene nanospheres and arch-like structures formed on the surface of the spheres during deposition.

### Acknowledgments

SEM images were acquired at the EMAL facility at the University of Michigan. We acknowledge DARPA Bio-Magnetics grant F008406 and NSF grant DMR 9900434 for financial support. Roy Clarke's group was supported in part by DoE grant DE-FG02-03ER46023.